\documentclass[pra,twocolumn,showpacs,superscriptaddress,floatfix]{revtex4}   

\font\cs=cmcsc10\font\sc=cmcsc10%
\font\tenmib=cmmib10 
\font\sevenmib=cmmib7
\font\fivemib=cmmib5 
\mathchardef\Ba   = "050B  
\mathchardef\Bb   = "050C  
\mathchardef\Bg   = "050D  
\mathchardef\Bd   = "050E  
\mathchardef\Be   = "0522  
\mathchardef\Bee  = "050F  
\mathchardef\Bz   = "0510  
\mathchardef\Bh   = "0511  
\mathchardef\Bthh = "0512  
\mathchardef\Bth  = "0523  
\mathchardef\Bi   = "0513  
\mathchardef\Bk   = "0514  
\mathchardef\Bl   = "0515  
\mathchardef\Bm   = "0516  
\mathchardef\Bn   = "0517  
\mathchardef\Bx   = "0518  
\mathchardef\Bom  = "0530  
\mathchardef\Bp   = "0519  
\mathchardef\Br   = "0525  
\mathchardef\Bro  = "051A  
\mathchardef\Bs   = "051B  
\mathchardef\Bsi  = "0526  
\mathchardef\Bt   = "051C  
\mathchardef\Bu   = "051D  
\mathchardef\Bf   = "0527  
\mathchardef\Bff  = "051E  
\mathchardef\Bch  = "051F  
\mathchardef\Bps  = "0520  
\mathchardef\Bo   = "0521  
\mathchardef\Bome = "0524  
\mathchardef\BG   = "0500  
\mathchardef\BD   = "0501  
\mathchardef\BTh  = "0502  
\mathchardef\BL   = "0503  
\mathchardef\BX   = "0504  
\mathchardef\BP   = "0505  
\mathchardef\BS   = "0506  
\mathchardef\BU   = "0507  
\mathchardef\BF   = "0508  
\mathchardef\BPs  = "0509  
\mathchardef\BO   = "050A  
\mathchardef\BDpr = "0540  
\mathchardef\Bstl = "053F  

\let\a=\alpha \let\b=\beta    \let\d=\delta 
\let\z=\zeta  \let\h=\eta   \let\th=\theta  
\let\m=\mu        \let\x=\xi     \let\p=\pi    \let\r=\rho
\let\s=\sigma \let\t=\tau   \let\f=\varphi 
  \let\ps=\psi   
\let\G=\Gamma \let\D=\Delta  \let\L=\Lambda 
    \let\Si=\Sigma     
\let\O=\Omega 

\def\*{\vglue0.3truecm}
\let\==\equiv
\let\0=\noindent
\def\ie{{\it i.e.\ }}
\def\rhs{{\it r.h.s.}\ }
\def\tende#1{\,\vtop{\ialign{##\cr\rightarrowfill\cr
 \noalign{\kern-1pt\nointerlineskip} \hskip3.pt${\scriptstyle
 #1}$\hskip3.pt\cr}}\,}
\def\otto{\,{\kern-1.truept\leftarrow\kern-5.truept\to\kern-1.truept}\,}
\def\media#1{{\langle#1\rangle}}
\def\defi{\,{\buildrel def\over=}\,}

\def\EE{{\cal E}}\def\NN{{\cal N}}\def\FF{{\cal F}}\def\CC{{\cal C}}
\newcommand\revtex{{R\kern-0.24mm\lower0.5mm\hbox{E}\kern-0.6mm V\kern-0.5mm%
\lower0.5mm\hbox{T}\kern-0.4mm E\kern-.2mm \lower0.5mm\hbox{X}}}
\def\V#1{{\bf#1}}
\def\lis#1{\overline#1}
\def\eg{{\it e.g.\ }}
\def\etc{{\it etc.\ }}
\def\ap{{\it a priori\ }}
\newdimen\xshift \newdimen\xwidth \newdimen\yshift \newdimen\ywidth

\def\ins#1#2#3{\vbox to0pt{\kern-#2\hbox{\kern#1 #3}\vss}\nointerlineskip}

\def\eqfig#1#2#3#4#5{
\par\xwidth=#1 \xshift=\hsize \advance\xshift
by-\xwidth \divide\xshift by 2
\yshift=#2 \divide\yshift by 2%
{\hglue\xshift \vbox to #2{\vfil
#3 \includegraphics{#4.eps}
}\hfill\raise\yshift\hbox{#5}}}

\font\tenmib=cmmib10 \font\eightmib=cmmib8
\font\sevenmib=cmmib7\font\fivemib=cmmib5 
\font\ottoit=cmti8\font\fiveit=cmti5\font\sixit=cmti6
\font\fivei=cmmi5\font\sixi=cmmi6\font\ottoi=cmmi8
\font\ottorm=cmr8
\font\ottosy=cmsy8\font\sixsy=cmsy6\font\fivesy=cmsy5
\font\ottobf=cmbx8\font\sixbf=cmbx6\font\fivebf=cmbx5%
\font\ottocss=cmcsc8%

\def\ottopunti{\def\rm{\fam0\ottorm}\def\it{\fam6\ottoit}%
\def\bf{\fam7\ottobf}%
\textfont1=\ottoi\scriptfont1=\sixi\scriptscriptfont1=\fivei%
\textfont2=\ottosy\scriptfont2=\sixsy\scriptscriptfont2=\fivesy%
\textfont4=\ottocss\scriptfont4=\sc\scriptscriptfont4=\sc%
\textfont5=\eightmib\scriptfont5=\sevenmib\scriptscriptfont5=\fivemib%
\textfont6=\ottoit\scriptfont6=\sixit\scriptscriptfont6=\fiveit%
\textfont7=\ottobf\scriptfont7=\sixbf\scriptscriptfont7=\fivebf%
\setbox\strutbox=\hbox{\vrule height7pt depth2pt width0pt}%
\normalbaselineskip=9pt\rm}
\let\nota=\ottopunti

\begin{document}

\voffset0.5truecm

\title{Entropy, Nonequilibrium, Chaos and Infinitesimals}
\*
\author{Giovanni Gallavotti}
\affiliation{Dipartimento di Fisica and INFN, Universit\`a di Roma 
{\em La Sapienza},
P.~A.~Moro 2, 00185, Roma, Italy}

\date{\today}

\begin{abstract} 
\it A survey of the approach to Statistical Mechanics
following Boltzmann's theory of ensembles and ergodic hypothesis
leading to chaoticity as a unifying principle of equilibrium and
nonequilibrium Statistical Mechanics.
\end{abstract}

\maketitle

\section{Boltzmann and Entropy}
\*

Since the earliest works {\cs Boltzmann} aimed at a microscopic
interpretation or ``proof'' of the newly formulated second Law of
Thermodynamics and of the associated concept of {\it entropy},
\cite{Bo66}.  At the time, since the works of {\cs Bernoulli,
Avogadro, Herapath, Waterstone, Kr\"onig, Clausius} it was well
established that there should be an identification between absolute
temperature and average kinetic energy at least in gas theory.

{\cs Boltzmann} starts in \cite{Bo66} by stating clearly that in
general (\ie not only for gases) temperature and average kinetic
energy must be identified and gives a derivation of the second law for
isocore (\ie constant volume) transformations. This first derivation
makes use of a periodicity assumption on the motion of (each) gas
particle to obtain the existence of the time average of the kinetic
energy and seems to fail if the motions are not periodic.

Nevertheless {\cs Boltzmann} insisted in conceiving aperiodic motions
as periodic with infinite period: see \cite{Bo66} where on p. 30 one
finds that ``{\it... this explanation is nothing else but the
mathematical formulation of the theorem according to which the paths
that do not close themselves in any finite time can be regarded as
closed in an infinite time}''. He therefore pursued the implied
mechanical proof of Thermodynamics to the extreme consequences.

A ``proof'' of the second law meant to look for properties of the
trajectories in phase space of a mechanical equation of motion, like
time averages of suitable observables, which could have the
interpretation of thermodynamic quantities like pressure $p$, volume
$V$, energy $U$, temperature $T$, entropy $S$ {\it and be related} by
the thermodynamic relations, namely

\begin{eqnarray}dS=\frac{dU\,+\,p\,dV}{T}\label{1.1}\end{eqnarray} 
where $dU,dV,dS$ are the variations of $U,V,S$ when the control parameters
of the system are infinitesimally changed.

The extreme consequence was the {\it ergodic hypothesis} which was
first mentioned by {\cs Boltzmann} around 1870, see p. 237 in
\cite{Bo871a}, for the internal motion of atoms in single molecules of
a gas (``{\it it is clear that the various gas molecules will go
through all possible states of motion}'' which, however, could
possibly be understood from the context to be different from the
ergodic hypothesis, see \cite{Br03}, because the molecules undergo
from time to time collisions). See also p. 96 in \cite{Bo868} and
p. xxxvii in the recent collection by {\cs Flamm} \cite{Fl00}.

Considering a collection of copies of the system alike to a large
molecule, p. 284 in \cite{Bo871b}, the same assumption became what is
often referred as the ergodicity property of the entire gas.  It
implied that, by considering all motions periodic, kinetic energy
equipartition would follow and, better (see p. 287 in \cite{Bo871b}),
even what we call now the microcanonical distribution would follow (as
well as the canonical distribution). The hypothesis was taken up also
by {\cs Maxwell} (1879), see p. 506 in \cite{Br03}).\footnote{[The
paper \cite{Bo871b} is a key work, albeit admittedly obscure: in
modern notations it considers a system of equations of motion with
dimension $n$ and $0$ divergence admitting $n-k$ constants of motion,
$\f_{k+1},\ldots,\f_n$, and are decribed by coordinates
$s_1,\ldots,s_n$. Then the distribution proportional to
$\prod_{j=k+1}^n \d(\f_j-a_j)\cdot \prod_{i=1}^n ds_i$ is {\it
invariant} and can be written ${\prod_{i=1}^k ds_i}\cdot\frac1{|\det
\partial(\f_{n-k+1},\ldots,\f_n)|}$ where the last denominator denotes the
Jacobian determinant (``last multiplier'') of $\f_{n-k+1},\ldots,\f_k$
with respect to $s_{k+1},\ldots,s_n$ evaluated at the given values
$a_{k+1},\ldots,a_n$ of the constants of motion ({\cs Boltzmann} calls
this an instance of the ``last multiplier principle'' of {\cs
Jacobi}). If the system has only one constant of motion, namely the
energy $H=\chi+\psi$ with $\chi\,=$ potential energy and $\ps\,=$
kinetic energy, this is the microcanonical distribution, as also
recognized by {\cs Gibbs} in the introduction of his book, \cite{Gi02}
(where he quotes \cite{Bo871b}, but giving to it the title of its
first section).]}

In this way {\cs Boltzmann} was able to derive various thermodynamic
consequences and a proof of \ref{1.1}, see \cite{Bo877a}, and was led to
exhibiting a remarkable example of what later would be called a
``{\it thermodynamic analogy}'' (Sec.III of \cite{Bo877a}). This meant the
existence of quantities associated with the phase space of a
mechanical equation of motion (typically defined as time averages over
the solutions of the equations of motion), which could be given
thermodynamic names like equilibrium state, pressure $p$, volume $V$,
energy $U$, temperature $T$, entropy $S$ {\it and be related} by the
thermodynamic relations that are expected to hold between the physical
quantities bearing the same name, namely \ref{1.1}.

The notion of mechanical thermodynamic analogy was formulated and
introduced by {\cs Helmoltz} for general systems admitting only
periodic motions (called {\it monocyclic}), \cite{He884a,He884b}.  The
proposal provided a new perspective and generated the new guiding idea
that the Thermodynamic relations would hold in {\it every mechanical
system}, from the small and simple to the large and complex: in the
first cases the relations would be trivial identities of no or little
interest, just {\it thermodynamic analogies}, but in the large systems
they would become nontrivial and interesting being relations of
general validity. In other words they would be a kind of symmetry
property of Hamiltonian Mechanics.

The case of spatially confined systems with one degree of freedom was
easy (easier than the example already given in \cite{Bo877a}): with all
motions periodic, the microscopic state was indentified with the phase
space point $(p,q)$ representing the full mechanical state of the
system, the {\it macroscopic state} in the corresponding thermodynamic
analogy was identified with the energy surface $H(p,q)=\frac1{2m}
p^2+\f_V(q)=U$, where $m$ is the mass and $\f_V$ is the potential energy
which confines the motion in position space, \ie in $q$, and depends
on a parameter $V$. The state is completely determined by two
parameters $U,V$.

Average kinetic energy $T=\lim_{\t\to\infty}\frac1\t\int_0^\t K(p(t)) dt$
is identified with temperature; energy is identified with $U$: then if
pressure $p$ is defined as the time average $\lim_{\t\to\infty}
-\frac1\t\int_0^\t \partial_V \f_V(q(t))dt$ the quantities $T,p$
become functions $p=p(U,V)$,$T=T(U,V)$ of the parameters $U,V$
determining the state of the system and the \ref{1.1} should hold.

Indeed the limits as $\t\to\infty$ exist in such a simple case, in which all
motions are periodic and confined between $q_\pm=q_\pm(U,V)$ (where
$U=\f_V(q_\pm)$); it is $dt\=\frac{dq}{|\dot
q|}=\frac{dq}{\sqrt{2(U-\f_V(q))/m}}$ and the period of the
oscillations is $\t_0=\t_0(U,V)=2\int_{q_-}^{q_+}
\frac{dq}{\sqrt{2(U-\f_V(q))/m}}$, hence (p. 127 in \cite{Bo884} and
Ch. I in \cite{Ga00}),

\begin{eqnarray}\label{1.2}
&T=\frac2{\t_0} \int_{q_-(U,V)}^{q_+(U,V)} \frac{m}{2}{
\sqrt{\frac 2m(U-\f_V(q))}}dq,
\qquad \\
&p=\frac2{\t_0} \int_{q_-(U,V)}^{q_+(U,V)} 
\frac{\partial_V\f_V(q)}{\sqrt{\frac m2(U-\f_V(q))}}dq\nonumber
\end{eqnarray}
and it is immediate to check, as in \cite{Bo877a}, that \ref{1.1} is
fulfilled by setting

\begin{eqnarray}\label{1.3}
&S(U,V)\,= \,2\, \log \int_{H=U} pdq=\nonumber\\
&=\,2\,\log \int_{q_-(U,V)}^{q_+(U,V)}
\sqrt{2m (U-\f_V(q))}\,dq\end{eqnarray}
The case of the central motion studied in \cite{Bo877a} was another
instance of {\it monocyclic} systems, \ie systems with only periodic
motions. 

Then in the fundamental paper \cite{Bo884}, following and
inspired by the quoted works of {\cs Helmoltz}, {\cs Boltzmann} was able to
achieve what I would call the completion of his program of deducing
the second law (\ref{1.1}) from Mechanics. If \*

\0(1) the {\it absolute temperature} $T$ is identified with the average
kinetic energy over the periodic motion following the initial datum
$(\V p,\V q)$ of a macroscopic collection of $N$ identical particles
interacting with a quite  {\it arbitrary} pair interaction, and

\0(2) the {\it energy} $U$ is $H(\V p,\V q)$ sum of kinetic and of a
potential energy,

\0(3) the {\it volume} $V$ is the volume of the region where the positions $\V
q$ are confined (typically by a hard wall potential),

\0(4) the {\it pressure} $p$ is the average force exercized on the walls by
    the colliding particles,
\*

\0then, from the assumption that each point would evolve periodically 
visiting every other point on the energy surface (\ie assuming that
the system could be regarded as monocyclic, see \cite{Ga00} Appendix
9.3 for details) it would follow that the quantity $p$ could be
identified with the $\media{-\partial_V\f_V}$, time average of $-\partial_V
\f_V$, and \ref{1.1} would follow as a {\it heat theorem}. The heat 
theorem would therefore be a consequence of the general properties of
monocyclic systems.

This led {\cs Boltzmann} to realize, in the same paper, that there
were a large number of mechanical models of Thermodynamics: the
macroscopic states could be identified with regions of phase spaces
invariant under time evolution and their points would contribute to
the average values of the quantities with thermodynamic interpretation
(\ie $p,V,U,T$) with a weight (hence a probability) also invariant
under time evolution.

Hence imagining the weights as a density function one would see the
evolution as a motion of phase space points leaving the density
fixed. Such distributions on phase space were called {\it monodic}
(because they keep their identity with time or, as we say, are
invariant): and in \cite{Bo884} several collections of weights or {\it
monodes} were introduced: today we call them collections of invariant
distributions on phase space or {\it ensembles}. Among the ensembles
$\EE$, \ie collections of monodes, {\cs Boltzmann} singled out the
ensembles called {\it orthodes} (``behaving correctly''): they were
the families of probability distributions depending on a few
parameters (normally $2$ for simple one component systems) such that
the corresponding averages $p,V,U,T$, defined in (1-4) above, would
vary when the parameters were varied causing variations $dU,dV$ of
average energy and volume in such a way that the \rhs of \ref{1.1}
would be an exact differential, thereby defining the {\it entropy} $S$
as a function of state, see \cite{Ga95a,Ga00}.

The ergodic hypothesis yields the ``orthodicity'' of the ensemble
$\EE$ that today we call {\it microcanonic} (in \cite{Bo884} it was
named {\it ergode}): but ergodicity, \ie the dynamical property that
evolution would make every phase space point visit every other, was
not necessary to the orthodicity proof of the ergode. In fact in
\cite{Bo884} the relation \ref{1.1} is proved directly without
recourse to dynamical properties (as we do today, see
\cite{Fi64,Ru68,Ga00}); and in the same way the orthodicity of the
{\it canonical ensemble} (called {\it holode} in \cite{Bo884}) was
obtained and shown to generate a Thermodynamics which is equivalent to
the one associated with the microcanonical ensemble.
\footnote{[Still today a different interpretation of the word
``ensemble'' is widely used: the above is based on what {\cs
Boltzmann} calls ``{\it Gattung von Monoden}'', see p.132, l. 14 of
\cite{Bo884}: unfortunately he is not really consistent in the use of
the name ``monode'' because, for instance in p. 134 of the same
reference, he clearly calls ``monode'' a collection of invariant
distributions rather than a single one; further confusion is generated
by a typo on p. 132, l. 22, where the word ``ergode'' is used instead
of ``holode'' while the ``ergode'' is defined only on p. 134. It seems
beyond doubt that ``holode'' and ``ergode'' were intended by Boltzmann
to be {\it collections} $\EE$ of invariant distributions
(parameterized respectively by $U,V$ or by $(k_B T)^{-1},V$ in modern
notations): Gibbs instead called ``ensemble'' each single invariant
distribution, or at least that is what is often stated. It seems that
the original names proposed by {\cs Boltzmann} are more appropriate,
but of course we must accept calling ``microcanonical ensemble'' the
ergode and ``canonical ensemble'' the holode, see \cite{Ga00}.]}

In the end in \cite{Bo871b} and, in final form, in \cite{Bo884} the
theory of ensembles and of their equivalence was born without need of
the ergodic property: the still important role of the ergodic
hypothesis was to guarantee that the quantities $p,V,U,T,S$ defined by
orthodic averages with respect to invariant distributions on phase
space had the physical meaning implied by their names (this was true
for the microcanonical ensemble by the ergodic hypothesis, and for the
other ensembles by the equivalence).

At the same time entropy had received a full microscopic
interpretation consistent with, but independent of, the one arising
from the {\it Boltzmann's equation} in the rarefied gases case, which
can be seen as a quite independent development of {\cs Boltzmann}'s
work. Furthermore it became clear that the entropy could be
identified, up to a universal proportionality constant $k_B$, with the
volume of phase space enclosed by the energy surface.

Unfortunately the paper \cite{Bo884} has been overlooked until quite
recently by many, actually by most, physicists possibly because it
starts, in discussing the thermodynamic analogy, by giving the Saturn
rings as an ``example'': a brilliant one, certainly but perhaps
discouraging for the suspicious readers of this deep and original
paper on Thermodynamics. See p.242 and p. 368 in \cite{Br76} for an
exception, possibly the first.

\section{Boltzmann's discrete vision of the ergodic problem}

The ergodic hypothesis could not possibly say that every point of the
energy surface in phase space visits in due time (the {\it recurrence
time}) every other, see also p. 505 and following in \cite{Br03}. But
this statement was attributed to {\cs Boltzmann} and criticized over
and over again (even by Physicists, including in the influential book,
\cite{EE11}, although enlightened mathematicians could see better, see
p.385 in \cite{Br76}): however for {\cs Boltzmann} phase space was
discrete and points in phase space were {\it cells} $\D$ with finite
size, that I will call $h$. And time evolution was a permutation of
the cells: ergodicity meant therefore that the permutation was a {\it
one cycle permutation}.

This conception, perfectly meaningful mathematically, was apparently
completely misunderstood by his critics: yet it was clearly stated
in one of the replies to {\cs Zermelo}, \cite{Bo96}, and in the book
on gases, \cite{Bo96a}, see also \cite{Ga95a} and the {\cs de
Courtenay}'s communication in this Symposium. 

In order to explain how a reversible dynamics could be compatible with
the irreversibility of macroscopic phenomena he had, in fact, to
estimate the recurrence time. This was done by multiplying the typical
time over which a microscopic event (\ie a collision) generates a
variation of the coordinates of an order of magnitude appreciable on
microscopic scales (\ie a time interval of $\sim10^{-12}$s and a
coordinate variation of the order of $1^o$A) times the number of cells
into which phase space was imagined to be subdivided.

The latter number was obtained by dividing the phase space around the
energy surface into equal boxes of a size $h$ equal to the $3N$-th
power of $\r^{-\frac13}$ times $\sqrt{m k_BT}$ with $\r$ the numerical
density and $k_B$ Boltzmann's constant and $T$ temperature. With the
data for $H_2$ at normal conditions in $1{\rm cm}^3$ an ealier
estimate of Thomson, \cite{Th74}, was rederived (and a recurrence time
scale so large that it would be immaterial to measure it in seconds or
in ages of the Universe).

Of course conceiving phase space as discrete is essential to formulate
the ergodicity property in an acceptable way: it does not, however,
make it easier to prove it even in the discrete sense just mentioned
(nor in the sense acquired later when it was formulated mathematically
for systems with continuous phase space). It is in fact very difficult
to be \ap\ sure that the dynamics is an evolution which has only one
cycle. Actually this is very doubtful: as one realizes if one
attempts a numerical simulation of an equation of motion
which generates motions which are ergodic in the mathematical sense.

And the difficulty is already manifest in the simpler problem of
simulating differential equations in a way which rigorously respects
the uniqueness theorem. In computers the microscopic states are
rigorously realized as cells (because points are described by
integers, so that the cells sizes are limited by the precision of
hardware and software) and phase space is finite. By construction
simulation programs map a cell into another: but it is extremely
difficult, and possible only in very special cases (among which the
only nontrivial that I know is \cite{LV93}) without dedicating an
inordinate computing time to insure a $1-1$ correspondence between the
cells.

Nevertheless the idea that phase space is discrete and motion is a
permutation of its points is very appealing because it gives a privileged
role to the {\it uniform distribution} on the phase space region in
which the motion develops (\ie the energy surface, if the ergodic
hypothesis holds). However it is necessary, for consistency, that the
phase space cells volume does not change with time, see Ch. 1 in
\cite{Ga00}: this is a property that holds for Hamiltonian evolutions
and therefore allows us to imagine the ergodic hypothesis as
consistent with the predictions of Statistical Mechanics.

\*
\section{Boltzmann's heritage}
\*

The success of the ergodic hypothesis has several aspects. One that
will not be considered further is that it is not necessary: this is
quite clear as in the end we want to find the relations between a very
limited number of observables and we do not need for that an
assumption which tells us the values of all possible averages, most of
which concern ``wild'' observables (like the position of a tagged
particle). The consequence is that the ergodic hypothesis is intended
in the sense that confined Hamiltonian systems ``can be regarded as
ergodic for the purpose of studying their equilibrium properties''.

What is, perhaps, the most interesting aspect of the hypothesis is that
it can hold for systems of any size and lead to relations which are
essentially size independent as well as model independent and which
become interesting properties when considered for macroscopic systems.

{\it Is it possible to follow the same path in studying nonequilibrium
phenomena?} The simplest such phenomena arise in stationary states of
systems subject to the action of nonconservative forces and of
suitable heat removing forces (whose function is to forbid indefinite
build up of energy in the system).

Such states are realized in Physics with great accuracy for very long
times, in most cases longer than the available observation times. For
instance it is possible to keep a current circulating in a wire
subject to an electromotive force for a very long time, provided a
suitable cooling device is attached to the wire.

As in equilibrium, the stationary states of a system will be described
by a collection of probability distributions on phase space $\EE$,
invariant with respect to the dynamics, which I call {\it
ensemble}: the distributions $\m$ in $\EE$ will be parameterized by a
few parameters $U,V,E_1,E_2,\ldots$ which have a physical
interpretation of (say) average energy, volume, intensity of the
nonconservative forces acting on the system (that will be called
``external parameters''). Each distribution $\m$ will describe a macroscopic
state in which the averages of the observables will be their integrals
with respect to $\m$.  The equations of motion will
be symbolically written as

\begin{eqnarray}\dot{\V x}=\V f(\V x)\label{3.1}\end{eqnarray} 
and we shall assume that $\V f$ is smooth, that it depends on the
external parameters and that the phase space visited by trajectories
is bounded (at fixed external parameters and initial data).

Since we imagine that the system is subject to nonconservative forces
the phase space volume (or any measure with density with respect to
the volume) will not be preserved by the evolution and the divergence

\begin{eqnarray}\s(\V x)=-\sum_{i}\partial_{x_i} f_i(\V x)\label{3.2}
\end{eqnarray}
will be {\it different} from $0$.

We expect that, in interesting cases, the time average $\s_+$ of
$\s$ will be positive:

\begin{eqnarray}\s_+\defi \lim_{\t\to\infty}\frac1\t \int_0^\t
\s(S_tx)\,dt\,>\,0\,.\label{3.3}\end{eqnarray} 
and, with few exceptions, $x$--independent.

This means that there cannnot be invariant distributions with density
with respect to the volume. And the problem to find even a single
invariant distribution is notrivial, except possibly for the ones
concentrated on periodic orbits.

The problem can be attacked, possibly, by following {\cs Boltzmann}'s
view of dynamics as discrete, (``{\it die Zahl der lebendigen Kr\"aft ist
eine diskrete}'', see p. 167 in \cite{Bo877b}).
\*

\section{Extending Boltzmann's ergodic hypothesis.}
\*

Consider a generic ``chaotic'' system described by equations like
\ref{3.1} which generate motions confined in phase space. Under very
general conditions it follows that $\s_+\ge0$, \cite{Ru96}, and we
concentrate on the case $\s_+>0$. The suggestion that phase space
should be regarded as discrete, and motion should simply be a one-cycle
permutation of the ``cells'' $\D$ representing the phase space points
is still very appealing as it would lead to the unambiguous
determination of the invariant distribution $\m$ describing the
statistical properties of the stationary states.

In fact this is an assumption implicit in any claim of physical
relevance of a simulation: as already mentioned above, a computer 
program defines a map on small cells in phase space. Already in the
case of Hamiltonian systems (\ie in equilibrium theory) a simulation
will not respect the uniqueness of solutions of the equation of motion
because the map between the cells will not be invertible: it is
extremely hard to write a program which avoids that two distinct cells
are mapped into the same cell (see above).

When $\s_+>0$ so that, in the average, phase space volume contracts the
uniqueness problem becomes essentially unsurmountable (and not only in
simulations); and there will be very many cells that eventually evolve into the
same cell: thus the evolution will not be a permutation of the
cells. It will, however, become {\it eventually} a permutation of a {\it
subset} of the initial set of cells. This reflects the fact that the
orbits of the solutions of the differential equation \ref{3.1} will
``cluster'' on an {\it attractor} which is a set of $0$ volume.

The conclusion is that the statistics of the motions will still be a
well defined probability distribution on phase space {\it provided}
the ergodic hypothesis is extended to mean that the permutation of the
cells on the attractor is a one-cycle permutation: it will be, in this
case, still the uniform distribution concentrated on the cells lying
on the attractor.

This viewpoint unifies the conception of the statistics of equilibrium
and of stationary nonequilibrium: the {\it statistics} $\m$ of the
motions, \ie the probability distribution $\m$ such that, in the
continuous version of the models,

\begin{eqnarray}\lim_{\t\to\infty}\frac1\t\int_0^\t F(S_tx)\,dt\,=
\,\int F(y) \,\m(dy)\,,\label{4.1}\end{eqnarray}
for all smooth observables $F$ and for all but a set of
zero volume of points $x$ on phase space, can be considered, {\it in
equilibrium as well as in stationary non equilibrium} states, as a
probability distribution which is uniform on the attractor.

The key obstacle to the above conception of Statistical Mechanics for
stationary states is that phase space cells cannot be supposed to
evolve, under the evolution assigned by \ref{3.1} when $\s_+>0$,
keeping a constant volume. Therefore regarding evolution as a map 
between cells of a discretized version of phase space contains new
sources of possible errors. Besides the error that is present in
equilibrium theory due to the cells deformations which leads to
violations of the uniqueness, \cite{Ga00}, there is an error due to
their contraction $\s_+>0$.

In equilibrium the first error can be reduced by reducing the cells
size and the time intervals at which the observations (to be
interpolated into the estimate of the integral in \ref{4.1}) are
taken. This is a nontrivial source of errors that can be estimated to
be physically acceptable, at least for the evaluation of the averages
of the few observables relevant for Thermodynamics, only in certain
regions of the phase diagrams, see Ch. I in \cite{Ga00}. But at least
in such regions the discrete interpretation of the ergodic hypothesis
leads us to a consistent representation of the evolution as a
permutation between discrete elements of a partition of phase space
into small cells.

Out of equilibrium the further source of discretization error due to
the actual reduction of phase space volume implies that it is not
consistent to view the motion as a permutation of cells of a
discretization of phase space into small equal volume elements.

A possible way out is to restrict attention to systems that show
strongly chaotic behavior. For instance systems which are transitive
(\ie admit a dense orbit) and hyperbolic, see \cite{GBG04} for a
formal definition, are typically chaotic systems which are also quite
well understood.

To enter into some detail it is convenient to look at the time
evolution by drawing a few surfaces $\Si_1,\Si_2,\ldots,\Si_s$
transversal to the phase space trajectories, and such that the
trajectories cross some of the surfaces over and over again (\ie each
trajectory crosses the surfaces infinitely many times both in the
future and in the past). Let $\Si=\cup_j \Si_j$ (usually called a
``Poincar\'e's section'') and let $S$ be the map
which transforms a point $\x\in \Si$ (\ie on one of the surface elements
$\Si_1,\Si_2,\ldots,\Si_s$) into the point $S\x$ where the orbit of
$\x$ meets again for the first time $\Si$ (\ie it is again on one of
the surface elements defining $\Si$). 

The points in phase space can therefore be described by pairs
$x=(\x,\th)$ if $\x$ is the point in $\Si$ last visited by the
trajectory starting at $x$ and $\th$ is the time elapsed since that
moment.

It is possible to partition $\Si$ into regions $P_1,P_2,\ldots,P_n$
with the property that the symbolic dynamics histories
$\Bs=\{\s_i\}_{i=-\infty}^\infty$ on the sets $P_\s$, $\s=1,\ldots,n$, has a
{\it Markov property}, in the sense that

(1) there is a suitable matrix $M_{\s,\s'}$ with entries $0$ or $1$,
such that if $M_{\s_i,\s_{i+1}}\=1$ for all $i$ then there is a unique
point $x$ such that $S^ix\in P_{\s_i}$: the point $x$ is said to be
``coded'' by the sequence $\Bs$. And

(2) calling {\it compatible} a sequence $\Bs$ with
$M_{\s_i,\s_{i+1}}\=1$ then for all points $x$ there is at least one
compatible sequence $\Bs$ which codes $x$ and for all but a set of
zero volume relative to $\Si$ the sequence $\Bs$ is unique (\ie much
as it is the case in the binary representation of real numbers).

The partition $P_1,P_2,\ldots,P_n$ is then called a {\it Markov
partition}: since the set of exceptions in the correspondence
$x\otto\Bs$ has zero volume, the volume distribution can be
represented as a probability distribution $\m_0$ over the space of
compatible sequences. And the statistics of the evolution of data $\x$
chosen at random with respect to the distribution $\m_0$, which is the
main object of interest, will therefore be represented also by a
$S$--invariant probability distribution on the space $\O$ of the
compatible sequences $\Bs$, \cite{GBG04}.

The sets $P_1,P_2,\ldots,P_n$ can be used to represent conveniently
the microscopic states of the system: given a precison $h>0$ it is
possible to find $N_h$ such that the sets 

\begin{eqnarray}P_{\s_{-N_h},\ldots,\s_{N_h}}=\bigcap_{j=-N_h}^{N_h} S^{-j}
P_{\s_j}\label{4.2}\end{eqnarray}
have a diameter $<h$. Therefore the (nonempty) sets
$\D=P_{\s_{-N_h},\ldots,\s_{N_h}}$ can be conveniently used as
``cells'' to describe the evolution, when the size $h$ is small enough
for considering acceptable to neglect the variations of the (few)
interesting observables within the $\D$'s.

The evolution $S$ will stretch $\D$ along the unstable planes and
compress it along the stable ones: it will map
$\D=P_{\s_{-N_h},\ldots,\s_{N_h}}$ inside the union of the $n$ sets
$\cup_\s P_{\s_{-N_h+1},\ldots,\s_{N_h},\s}$. 

We then imagine that the
cell $\D$ is filled by smaller boxes, that will be called {\it
microscopic cells} or simply {\it microcells}, of equal volume, which
under the action of $S$ are transformed into boxes contained in only
one of the $n$ sets $\D_{\s}=P_{\s_{-N_h+1},\ldots,\s_{N_h},\s}$. The microcells,
which in a simulation could be identified with the integers defining
them in the computer memory, should be thought of as arranged in layers
adjacent to unstable planes of $S$ and are mapped into
microcells of the corresponding layers in the $n$ cells $\D_\s$.

Since the evolution, in the average, contracts phase space the layers
will merge under the action of $S$ so that the number of microcells
will initially decrease; but eventually in each cell $\D$ will survive
layers of microcells whose collection will be mapped one to one into
itself: the latter collection of microcells is a representation of the
attractor within the precision $h$. This is illustrated symbolically
in Fig.1.

\eqfig{155pt}{32pt}{}{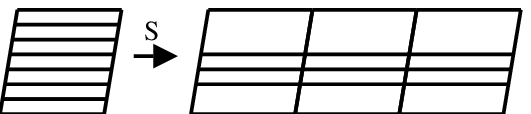}{}

\0{\nota Fig.1: The lines symbolize arrays of microcells $\D$: the ones
in the left drawing are stretched and merged by the time evolution
into arrays that end up in three new boxes.\vfil}

For consistency the number of microcells that is eventually found in each
cell $\D$ is inversely proportional to the expansion rate
$\L_e(\D)^{-1}$ of the surface elements on the unstable manifold in
$\D$: it will be denoted $\NN\L_e(\D)^{-1}$

The time evolution $S$ can then be represented as a permutation of the
$\NN\sum_\D\L_e(\D)^{-1}$ microcells on the attractor:

\* 
\0(1) give a rule to select the $\NN\L_e(\D)^{-1}$ microcells in
$\D=P_{\s_{-N_h},\ldots,\s_{N_h} }$ and to partition them into $n$ groups
labeled by $\s=1,\ldots,n$, each containing a fraction
$\p_\s=\frac{\L_e^{-1}(\D_{\s})}{\sum_{\s'} \L_e^{-1}(\D_{\s'})}$ of
the microcells selected in $\D$ and 
\\ 
(2) establish a correspondence $\lis S$ between the microcells in the
group labeled $\s$ and a subset the ones in
$\D_{\s}=P_{\s_{-N_h+1},\ldots,\s_{N_h},\s}$ 
\\ 
(3) approximate $S$ by replacing it by $\lis S$.  \*

Certainly there is a lot of ambiguity in deciding how to set up the
selection and the correspondence: but for the purposes of a
description of dynamics with precision $h$ the ambiguity has no
consequence. Note that in simulations the microcells selection is
implicitly prescribed by the program, and certainly changes quite
substantially by any small change of the program or by a change of the
computer used. By definition of program the evolution $S$ is replaced
by a map of microcells (in huge number even in simple simulations):
the map is not invertible but being a map of a finite set into itself
it is eventually reduced to a permutation of a {\it subset} of the
microcells.

Transitivity of the compatibility matrix $M$ implies that the
permutation of the microcells on the attractor can be chosen cyclic:
therefore the stationary distribution $\m$ will be approximated by the
uniform distribution on the attractor: this is a picture which seems
close to {\cs Boltzmann}'s conception of discretness and extends the
ideas behind the ergodic hypothesis to more general dynamical systems,
\cite{Ga96,Ga00}.  \*

\section{A bit of history.}
\*

The discovery of the probability distribution $\m$ that describes the
statistics of the stationary states of dynamical systems with confined
evolution did not follow the path discussed in Sec.4: of course
every theorem is preceded by a heuristic intuition and the exact
genesis of the ideas should be asked to their Authors at least to the
ones present here. But there is no certainty that they will give a
faithful account, as it is well known that recollection of past events in
the human mind, even important ones, tends to be modified as years
pass and new events occur.

A possible history about {\cs Boltzmann}'s ergodic hypothesis and theory of
ensembles is presented in Sec.1,2. 

The theory of hyperbolic transitive systems is much more recent: they
were formally introduced by {\cs Anosov} who proved the stability of
the notion under perturbations: a hyperbolic transitive system remains
such if slightly perturbed.

The existence of a well defined statistics for almost all initial data
was heralded, \cite{AW70}, by the work of {\cs Adler} and {\cs Weiss}
on the area preserving map of the torus
$S(\f_1,\f_2)=(\f_1+\f_2,\f_1+2\f_2)$: they define and make use of a
Markov partition. The idea was independently developed by {\cs Sinai},
\cite{Si68a,GBG04}, who treated the general case of an Anosov map, building
Markov partitions and using them to prove the existence of a
privileged distribution $\m$ giving the statistics of all initial data
but a set of zero volume.

A remarkable property of the distribution $\m$ emerges when it is
regarded as a probability distribution on the compatible sequences
$\Bs$ which code the points $\V x$ phase space. Namely it is a ``Gibbs
distribution'', in the sense of probability theory, with a short range
potential: this is, essentially, a Markov process with finitely many
states, \ie an object that is very well understood, \cite{Si68b}. The
surprising consequence is that Anosov systems are ``completely
integrable'' in the sense that we can compute essentially everything
at least in principle, \cite{Si72,GBG04}. They become a {\it paradigm} for
chaotic evolutions in the same sense in which harmonic oscillators are
a paradigm for ordered motions.

Hyperbolicity is a strong property which in practice is not satisfied
in physical systems. Therefore {\cs Bowen}, \cite{Bo70a}, and {\cs
Ruelle}, \cite{BR75,Ru76}, developed a theory for systems that are
hyperbolic in a much weaker sense: these are systems with {\it axiom
A} attractors. Also for such systems it is possible to define a
natural distribution that describes the statistics of all but a set of
zero volume of initial data and the distribution can be studied by an
extension of {\cs Sinai}'s methods.

The natural distribution has since been called the {\it SRB
distribution} and {\cs Ruelle} has proposed, in the very similar
context of turbulence theory, and at least three years earlier than it
appeared in print \cite{Ru76,Ru80}, that in general there should be a
unique distribution (or possibly a finite number of them) describing
the experimental statistics of motions: it should be the distribution
giving the asymptotic behavior of motions with arbitrary initial data
apart from a set of zero volume.  \*

\section{Developments. }
\*

More recently {\cs Cohen} and myself, \cite{GC95} and see also
\cite{Ga00}, proposed cutting a ``Gordian node'' by an hypothesis
which extends the viewpoint expressed by {\cs Ruelle} in \cite{Ru73}
``{\it ... while one would be very happy to prove ergodicity because it
would justify the use of Gibbs' microcanonical ensemble, real systems
perhaps are not ergodic but behave nevertheless in much the same way
and are well described by Gibbs' ensemble...}'': 
\*

{\bf Chaotic hypothesis:} {\it The asymptotic motions of a confined
chaotic mechanical system develop on an attracting set on which motion
can be considered a mixing Anosov flow.}  \*

Of course this applies in particular to Hamiltonian systems (where the
attracting set is the full energy surface) and it implies ergodicity:
hence the whole body of equilibrium Statistical Mechanics; {\it
furthermore} it puts on the same level equilibrium and non
equilibrium.
 
This interpretation of {\cs Ruelle}'s ideas, \cite{Ru73,Ru80}, has some
applications because it implies a {\it formal} expression of the
average values of the observables. Even though the expression is not
(yet ?) computable in any interesting case it may be useful to
establish relations between average values. For instance implications
of a microscopic symmetry on macroscopic observables might be found
from the formal (even if not practically computable) expression of the
SRB distribution.

An example is obtained by considering an Anosov system $(\FF,S)$, with
$\FF$ a smooth bounded manifold and $S$ a smooth transitive hyperbolic
map of $\FF$. Let 

\begin{eqnarray}\s(x)=-\log|\det \partial_x S(x)|\label{6.1}\end{eqnarray} 
be the phase space contraction rate and let $\m$ be the SRB distribution;
suppose 
\\(a) {\it dissipativity}, \ie $\s_+=\int \s(y)\m(dy)>0$, and
\\(b) {\it time reversal symmetry} in the sense that there is a smooth
isometry $I$ such that $IS=S^{-1}I$.
\\Define

\begin{eqnarray}p=\frac1\t\sum_{j=0}^{\t-1}
  \frac{\s(S^jx)}{\s_+}\label{6.2}
\end{eqnarray}
then the following theorem holds, \cite{GC95},
\*

{\bf Fluctuation theorem:} {\it With respect to the SRB distribution
the observable $p$ satisfies a large deviation property ({\rm see
below}) with a rate function $\z(p)$ which is analytic and convex in
an interval $(-p^*,p^*)$, for a suitable $p^*\ge 1$, where it exhibits
the symmetry property

\begin{eqnarray}\z(-p)=\z(p)-p\s_+\label{6.3}\end{eqnarray}
}
\*

This means that the probability, with respect to the SRB distribution
$\m$ of $(\FF,S)$, that $p$ is inside an interval $[a,b]\subset
(-p^*,p^*)$ is $P_{a,b}$ with $\lim_{\t\to\infty} \frac1\t\log
P_{a,b}=\max_{p\in[a,b]} \z(p)$.

Existence and analyticity of $\z(p)$ is part of the quoted general
results of {\cs Sinai}, while the symmetry \ref{6.3} was pointed out
in \cite{GC95} in an attempt to explain the numerical results of an
earlier computer experiment \cite{ECM93}. The interest of the theorem
lies in the fact that it is a symmetry property: hence it holds
without any free parameter.

The theorem can be extended to mixing Anosov flows, \cite{Ge98}, and
therefore, via the chaotic hypothesis and if $\s(x)$ is the phase
space contraction rate defined in \ref{3.2}, it becomes a property of
essentially any system which is {\it chaotic, dissipative and
reversible}.
\*

\section{Entropy?}
\*

Interest in the properties of the observable $\s(x)$, \ref{3.2}
for flows and \ref{6.2} for maps, arose in several molecular dynamics
simulations in which it was naturally related to the {\it entropy
creation rate}.

A natural question is whether a definition of entropy can be extended
to nonequilibrium stationary states in analogy with the
corresponding definition for equilibrium states (which are a very
special case of stationary states).

The identification between the SRB distributions and distributions
giving equal probability to the microcells in the attractor shows that
it should be possible, at least, to define a function which is a
Lyapunov function for the approach to stationarity: this would be an
extension of the $H$-theorem of Boltzmann. However equality between
the $H$ function evaluated in equilibrium states and thermodynamic
entropy might be a coincidence, important but not extendible to non
equilibrium (hence not necessary). Arguments in this direction can be
found in the literature, \cite{Ga04b,Ga01}, and here the controversial
aspects of this matter will not be touched, \cite{GL03}.

It will be worth however to enter into more details about why $\s(x)$
has been called entropy creation rate. This is simply because in
several experiments it had such an interpretation, being the ratio
between a quantity that could be identified with the work per unit
time done by the noncoservative forces stirring the system divided by a
quantity identified with temperature of the thermostat providing the
forces that extract the energy input from the stirring forces. The
experiments were simulations and from many sides critiques were
expressed because the interpretation seemed closely tied to the
explicit form of the thermostats models, often considered
``unphysical''.

Furthermore the explicit dependence on the equations of motion makes
the identification of $\s(x)$ with the entropy creation rate quite
useless if the aim is to compare the theory with experiments different
from simulations because in real experiments (\ie on experiments on
matter distinct from impressive arrays of transistors) there usually
is no explicit model of thermostat force and it is difficult to
evaluate $\s(x)$. And it might turn out that the identification of
$\s(x)$ with entropy creation rate is closely related to the special
models considered.

A simple, but quite general, model of thermostatted system may be
useful to show that, while we should expect that there is a relation
between entropy creation rate and phase space contraction, still the
two notions are quite different.

The system consists in $N\equiv N_0$ particles in a container $\CC_0$
and of $ N_a$ particles in $n$ containers $\CC_a$ which play the role
of {\it thermostats}: their positions will be
denoted $\V X_a,\,a=0,1,\ldots,n$, and $\V X\defi(\V X_0,\V
X_1,\ldots,\V X_n)$.  Interactions will be described by a potential
energy

\begin{eqnarray}
W(\V X)=\sum_{a=0}^{n} U_a(\V X_a) +\sum_{a=1}^n W_a(\V X_0,\V X_a)
\label{7.1}\end{eqnarray}
{\it i.e.} thermostats particles only interact indirectly, 
via the system. All masses will be $m=1$, for
simplicity.

The particles in $\CC_0$ will also be subject to external, possibly
nonconservative, forces $\V F(\V X_0,\BF)$ depending on a few strength
parameters $\BF=(E_1,E_2,\ldots)$. It is convenient to imagine that
the force due to the confining potential determining the region
$\CC_0$ is included in $\V F$, so that one of the parameters is the
volume $V=|\CC_0|$. See Fig.2 below.

\eqfig{110pt}{90pt}{}{fig2}{}

\0{\nota Fig.2 The reservoirs occupy finite regions outside $C_0$, \eg sectors
$C_a\subset R^3$, $a=1,2\ldots$. Their particles are constrained to
have a {\it total} kinetic energy $K_a$ constant, by suitable forces
$\Bth_a$, so that the reservoirs ``temperatures'' $T_a$, see
\ref{7.3}, are well defined.\vfil}  \*

\kern-3mm 
The equations of motion will be, assuming the mass $m=1$,

\begin{eqnarray}\label{7.2}
&\ddot{\V X}_{0i}=-\partial_i U_0(\V X_0)-\sum_{a}
\partial_i U_a(\V X_0,\V X_i)+\V F_i\\
&\ddot{\V X}_{ai}=-\partial_i U_a(\V X_a)-
\partial_i U_a(\V X_0,\V X_i)-\a_a \dot{\V X}_a\nonumber\end{eqnarray}
where the last force term $-\a_a \dot{\V X}_a$ is a phenomenological
force that implies that the thermostats particles keep constant
kinetic energies:

\begin{eqnarray}K_a=\sum_{j=1}^{N_a} \frac12\, (\dot{\V X}^a_j)^2\defi
\frac32 N_a k_B T_a\defi \frac32 N_a\b_a^{-1}\label{7.3}\end{eqnarray}
where the parameters $T_a$ should define the thermostats {\it
temperatures} and $\a_a$ can, for instance, be defined by

\begin{eqnarray}-\a_a \,\defi\,\frac{L_a-\dot U_a} {3N_a k_B
    T_a}\label{7.4}
\end{eqnarray} 
where $L_a=-\partial_{\V X_a} W_a(\V X_0,\V X_a)\cdot \dot{\V X}_a$ is the
work done per unit time by the forces that the particles in $\CC_0$
exert on the particles in $\CC_a$.

The exact form of the forces that have to be added in order
to insure the kinetic energies constancy should not really matter,
within wide limits. But this is a property that is not obvious and
which is much debated. The above thermostatting forces choice is
dictated by Gauss' {\it least effort} principle for the constraints
$K_a=const$, see appendix 9.4 in \cite{Ga00}: this is a criterion that
has been adopted in several simulations, \cite{EM90}. Independently of
Gauss' principle it is immediate to check that if $\a_a$ is defined by
\ref{7.4} then the kinetic energies $K_a$ are strictly constants of
motion.

The work $L_a$ in \ref{7.4} will be interpreted as {\it heat} $\dot
Q_a$ ceded, per unit time, by the particles in $\CC_0$ to the $a$-th
thermostat (because the ``temperature'' of $\CC_a$ remains constant,
hence the thermostats can be regarded in thermal equilibrium).  The
{\it entropy creation rate} due to heat exchanges between the system
and the thermostats can, therefore, be naturally defined as

\begin{eqnarray}\s^0(\dot{\V X},\V X)\defi\sum_{a=1}^{N_a} \frac{\dot Q_a}{k_B
T_a}\label{7.5}\end{eqnarray} 
It should be stressed that here {\it no entropy notion} is introduced
for the stationary state: only variation of the thermostats entropy is
considered and it should not be regarded as a new quantity because the
thermostats should be considered in equilibrium at a fixed
temperature.

The question is whether there is any relation between $\s_0$ and the
phase space contraction $\s$ of \ref{3.2}. The latter can be
immediately computed and is (neglecting $O(\min_{a>0} N_a^{-1})$)

\begin{eqnarray}\s^\G(\dot{\V X},\V X)={\mathop\sum\limits_{a>0}}
{\frac{3N_a-1}{3 N_a}} \frac{\dot Q_a-\dot U_a}{k_B T_a} =
{\mathop\sum\limits_{a>0}} \frac{\dot Q_a}{k_B T_a}-\dot U\label{7.6}
\end{eqnarray} 
where $U=\sum_{a>0} \frac{3N_a-1}{3 N_a} \frac{U_a}{k_B T_a}$.  Hence in
this example in which the thermostats are ``external''
to the system volume (unlike to what happens in the common examples in
which they act inside the volume of the system), the phase space
contraction is not the entropy creation rate, \cite{Ga06}.  {\it
However  it differs from the entropy creation rate by a total derivative}.

The latter remark implies that if the chaotic hypothesis is accepted
for the system in Fig.2 then, assuming $U_a$ bounded (for simplicity,
see \cite{BGGZ05,Ga06} for more general cases) it is
$\s_+=\media{\s_0}$ because the derivative $\dot U$ contributes
$\frac1\t(U(\t)-U(0))\tende{\t\to\infty}0$ and also the observable $p$, in
the continuous time extension of \ref{6.2}, \cite{Ge98}, has the same
rate function as the observable $p=\frac1\t\int_0^\t
\s_0(S_tx)\,dt\=\frac1\t\int_0^\t \s(S_tx)\,dt+O(\t^{-1})$. Since the
equations of motion \ref{7.2} are time reversible (a rather general
property of Gaussian constraints, with $I$ being here simply velocity
reversal) it follows that {\it the ``physical entropy creation''
\ref{7.5} has a fluctuations rate $\z(p)$ satisfying the fluctuation relation
\ref{6.3}.}

This is relevant because the definition \ref{7.5} has meaning
independently of the equations of motions and can therefore be suitable
for experimental tests. \cite{BGGZ06,Ga06}. The above is just a model
of thermostats: other interesting models have been proposed based on
purely Hamiltonian interactions at the price of relying on thermostats
of infinite size, see \cite{Ja99,EPR99,Ru06}.
\*

\section{Extensions of Boltzmann's $H$-theorem}
\*

The above analysis {\it does not require a notion of entropy} to be defined
for stationary states. 

There is, however, another key contribution of Boltzmann to
Statistical Mechanics, briefly mentioned above. This is the
Boltzmann's equation and the relative {\it H-theorem}, \cite{Bo72}.

The theorem has attracted deep interest because of its philosophical
implications. For our purposes it is important because it provides a
theory of approach to equilibrium and therefore it is one of the first
results on nonequilibrium. 

It is useful to stress that the definition of $H$ is given in the
context of the approach to equilibrium and {\cs Boltzmann} never
applied it (nor, perhaps, meant to apply it) to the approach to other
stationary states and to their theory. The equality of the value of
$H$ with the thermodynamic entropy when evaluated on the equilibrium
state raised the hope that it could be possible to define entropy for
systems out of equilibrium and even if not in stationary state. The
idea emerged clearly already from the foundational papers on the
Boltzmann equation (``{\it $\log P$ was well defined whether or not
the system is in equilibrium, so that it could serve as a suitable
generalization of entropy}'', p. 82 in \cite{Kl73} and p. 218 in
\cite{Bo877b}) and many attempts can be found in the literature to
define entropy for systems out of equilibrium in stationary states or
even in macroscopically evolving states.

Strictly speaking the implication that can be drawn from the works on
the Boltzmann's equation is that a rarefied gas started in a given
configuration evolves in time so that the average values of the
observables, at least of the few of interest, acquire an asymptotic
value which is the same as the one that can be computed from a
probability distribution maximizing a function $H$.

The acquisition of an asymptotic value by the averages of the
observables is a property expected to hold also when the
asymptotic state is a nonequilibrium stationary state. And it is natural to
think that also in such cases there will be a function that approaches
monotonically an asymptotic value signaling that the few observables
of interest approach their asynptotic average.

As remarked above the SRB distribution is a uniform distribution over
the attractor: therefore it verifies a variational property and this
can be used to define a Lyapunov function that evolves towards a
maximum, \cite{Ga04b}. Let $\Bh=(\s_{-N_h},\ldots,\s_{N_h})$ and
$H\defi\frac1\t\sum_{\Bh} -p_{\Bh}\big(\log p_{\Bh}\ +\log
\L_\t(\Bh^{-1})\big)$ where $p_{\Bh}$ denotes the fraction of
microcells that can be found in the cell
$\D=P_{\Bh}=\cap_{k=-N_h}^{N_h}S^{-k} P_{\h_k}$ after a time of $\t$ units
has elapsed starting from an initial distribution $p^0_\Bh$ (typically
a uniform distribution over the microcells in a single cell
$\D^0$). This is a quantity that tends to a maximum as time evolves
(reaching it when the $ p_\Bh$ have the value of the SRB distribution
and the maximum equals, therefore, the logarithm of the number of
microcells on the attractor).

Therefore the quantity $H$ tends in the average to a maximum and it
can be regarded as an instance of an $H$--function. However the
maximum depends on the precision $h$ of the coarse graining defined by
the partition of phase space by the cells $\D$. Changing the precison
several changes occur which have to be examined if a meaning other
than that of a Lyapunov function has to be given to $H$. The analysis
in \cite{Ga04b} points out that $H$ changes with the precision $h$ in
a trivial way (\ie by an additive constant, independent of the control
parameters of the system and depending only on the precision $h$) if
the SRB state on which it is evaluated is an equilibrium state. In the
latter case it is proportional to the logarithm of the phase space
volume that can be visited. In the nonequilibrium cases however $H$
changes when the precision $h$ changes by additive quantities that
{\it are not just functions of $h$} but depend on thermodynamic
quantities, (like average energy, temperatures, {\etc.}),
\cite{Ga04b}.

This indicates that while not excluding the possibility of existence
of Lyapunov functions, see \cite{GGL04}, indicating the approach to
equilibrium (within a given precision $h$) the identity of the $H$
function with entropy, \ie its identity with a function of the state
parameters of the system, is possible only when the state is in an
equilibrium state. My interpretation of this analysis, based once more
on a discrete point of view on the problem, is that one should not
insist in looking for a notion of entropy in systems out of
equilibrium, \cite{Ga04b}.

If so once again {\cs Boltzmann}'s attitude to consider phase space as
discrete and in general to deny reality to the continua might have led
to insights into difficult questions.

\*
\section{Conclusion}
\*

{\cs Boltzmann}'s contribution to the theory of ensembles and to
the mechanical interpretation of heat and Thermodynamics was based on
a discrete conception of the continuum: his staunch coherence on this
view has been an essential aspect of the originality of his thought.

It is in fact a method of investigation which is still very fruitful
and used in various forms when ``cut--offs'' or ``regularizations''
are employed in the most diverse fields. In my view it has been and
still is important in the recent developments in the theory of
nonequilibrium stationary states. The Fluctuation theorem and its
various interpretations, extensions and applications (to Onsager
reciprocity at non zero forcing, to Green-Kubo formulae, to fluid
Mechanics, Turbulence and Intermittency, see \cite{Ga00,Ga04,GBG04}) is,
hopefully, only an example.

It is interesting in this context recall a few quotes from {\cs Boltzmann}
\*

``{\it Through the symbols manipulations of integral calculus, which have
become common practice, one can temporarily forget the need to start
from a finite number of elements, that is at the basis of the creation
of the concept, but one cannot avoid it}'';

\0see p. 227 in \cite{Bo874}, or in the same page:

``{\it Differential equations require, just as atomism does, an
initial idea of a large finite number of numerical values and points
...... Only afterwards it is maintained that the picture never
represents phenomena exactly but merely approximates them more and
more the greater the number of these points and the smaller the
distance between them. Yet here again it seems to me that so far we
cannot exclude the possibility that for a certain very large number of
points the picture will best represent phenomena and that for greater
numbers it will become again less accurate, so that atoms do exist in
large but finite number.}'' 

\0and, see p. 55 in \cite{Bo874}:

``{\it This naturally does not exclude that, after we got used once and
for all to the abstraction of the volume elements and of the other
symbols {\rm[of Calculus]} and once one has studied the way to operate
with them, it could look handy and luring, in deriving certain
formulae that Volkmann calls formulae for the coarse phenomena, to
forget completely the atomistic significance of such
abstractions. They provide a general model for all cases in which one
can think to deal with $10^{10}$ or $10^{10^{10}}$ elements in a cubic
millimeter or even with billions of times more; hence they are
particularly invaluable in the frame of Geometry, which must equally
well adapt to deal with the most diverse physical cases in which the
number of the elements can be widely different. Often in the use of
all such models, created in this way, it is necessary to
put aside the basic concept, from which they have overgrown, and perhaps
to forget it entirely, at least temporarily. But I think that it would
be a mistake to think that one could become free of it
entirely.}''
\*

And the principle was really applied not only in the conception of the
ergodic hypothesis, \cite{Bo871a,Bo871b}, but also in the deduction of
the Boltzmann's equation which {\cs Boltzmann} felt would be clarified
by following discretization methods (in energy) inspired by those
employed in the ``{\it elegant solution of the problem of
string-vibrations}'' of {\cs Lagrange}, or in {\cs Stefan}'s study of
diffusion or in {\cs Riemann}'s theory of mean curvature, \cite{Bo72}
and in various discussions of the heat theorem, \cite{Bo877b}.

The above conception of the infinitesimal quantities, rooted in the
early days of Calculus when ``$dx$'' was regarded as {\it infinitely
small and yet still of finite size} (in apparent, familiar, logical
contradiction), is an important legacy that should not be forgotten in
spite of the social pressure that induces all of us to identify
clarity of physical understanding with continuous models of reality.
\*

\0{Appendix: \it Temperature and kinetic energy, \cite{Br03,Br76}}
\*

The first attempts at a kinetic explanation of the properties of gases
came following the experiments by {\cs Boyle}, (1660), on the gas
compression laws.  The laws established that ``air'' had elastic
properties and that there was inverse proportionality between pressure
and volume: a theory that was considered also by {\cs Newton}. It was
{\cs D. Bernoulli}, (1720), who abandoned the view, espoused by {\cs
Newton}, that the atoms were arranged on a kind of lattice repelling
each other (with a force inversely proportional to their distances to
agree with Boyle's law, but extending only to the nearest
neighbors). {\cs Bernoulli} imagined the atoms to be free and that
pressure was due to the collisions with the walls and proportional to
the square of the average speed proposing that a correct definition of
temperature should be based on this property.

In 1816 {\cs Avogadro} established that, for rarefied gases, the ratio
$pV/T$ is proportional to the number of atoms or molecules via a
universal constant. This was a striking result, explaining the
anomalies in the earlier theory of {\cs Dalton} and allowing, besides
the definition of the {\it Avogadro's number}, the correct
determination of the relative molecular and atomic weights. It openend
the way to the definition of absolute temperature, independently of
the special gas-thermometer employed, and to the principle of energy
equipartition and to the later works of {\cs Waterston, Clausius,
Boltzmann}, among others.

The attempt of {\cs Laplace}, (1821), proposed an elaborate scheme in
which the atoms, still essentially fixed in space at average distance
$r$ would contain a quantity $c$ of {\it caloric} and would interact
with a short range force proportional to the product of their quantity
of caloric and depending on the distance. Identifying the caloric $c$
with the a fixed amount contained in each atom would have led to a gas
law with $p$ proportional to the square of the density $\r$, \ie to
$\r^2 c^2$; but this was avoided by supposing that the amount of
caloric $c$ in each molecule was determined by an equilibrium between
the amount of caloric emitted by a molecule and the caloric received
by it (emitted from the other molecules) which was supposed to depend
only on the temperature, see \cite{Br03}.

The theory of {\cs Laplace} did not sound convincing and the work of
{\cs Bernoulli} went unnoticed; the same was the fate of the work of
{\cs Herapath}, (1820), who again proposed, without knowing
{\cs Bernoulli}'s theory, that the atoms were free and pressure was
due to collisions with the walls; however he assumed that pressure
was proportional to the average momentum rather than kinetic energy
obtaining an incorrect definition of absolute temperature. In any
event his work was rejected by the {\it Philosophical transactions of
the Royal Society of London} and published on the {\it Annals of
Philosophy} falling into oblivion for a while.

In 1845 {\cs Waterstone}, unaware of both {\cs Bernoulli} and {\cs
Herapath} but (likely) familiar with {\cs Avogadro}'s work, proposed
the theory of gases with the correct identification of pressure as
proportional to the average kinetic energy and the density,
introducing also a rather detailed conception of te interatomic forces
taking up ideas inspired by {\cs Mossotti} (who probably had also made
{\cs Avogadro} and Italian science better known in England during his
political exile). Unfortunately he submitted it to the {\it
Philosophical transactions of the Royal Society} which readily
rejected it and remained unpublished, until it was rediscovered much
later (1892, by {\cs Raileigh}).

In the 1840's, through the work of {\cs Meyer, Joule, Helmoltz} and others
the energy conservation principle was established with the consequent
identification of heat as a form of energy convertible into mechanical
work forcing (reasonable) physicists to abandon the hypothesis of the
existence of caloric as a conserved entity.

The theory of gases begun to be really accepted with the work of {\cs
Kr\"onig}, (1856), who clearly proposed identifying temperature with
average kinetic energy of molecules. His work became well known as it
appeared to have prompted the publication of {\cs Clausius}'s paper of
(1857), who had independently reached the same conclusions and gone
much further. Not only {\cs Clausius} went quite far in establishing
energy equipartition (completed by {\cs Maxwell} in 1860) but he
introduced a basic concept of kinetic theory: the mean free path. Thus
making clear the role of collisions in gas theory: they lead to
prediction and to a first understanding of the phenomenon of diffusion,
explaining the apparent paradoxes linked to the earlier assumptions
that in rarefied gases collisions could be simply neglected, and also
initiate the theory of the transport coefficients.

The latter papers, one century after the too far in advance (over his
time) work of {\cs Bernoulli}, gave origin to kinetic theory in the
sense we intend it still now, and stimulated also the related
investigations of {\cs Maxwell}. Therefore {\cs Maxwell} (1859) and a
little later {\cs Boltzmann} (1866) could start their work taking for
granted the well established identity between temperature and average
kinetic energy for gases extending it to hold in all systems in
equilibrium (rarefied or not). This key view was not destined to have
a long life: the advent of Quantum Mechanics would prove that
proportionality between average kinetic energy and temperature could
only be approximate and to hold if quantum corrections to Atomic
Mechanics were negligible, see \cite{Ga00}.  Nevertheless the
identification of temperature and kinetic energy plaid (and still
plays, whenever quantum effects are negligible) an essential role not
only in classical Statistical Mechanics but also in the discovery of
Quantum Mechanics, which was heralded by the failure of the related
equipartition of energy.

\*
\0{\it Source of the talk at the {\cs Boltzmann's Legacy}
international symposium at ESI, Vienna, 7-9 June, 2006}

\nota
\bibliographystyle{apsrev}

\revtex
\end{document}